\documentclass[a4paper,12pt]{article}
\usepackage{amssymb,latexsym,amsmath}

\makeatletter \@addtoreset{equation}{section} \makeatother

\input epsf

\usepackage{pdfsync}
\usepackage{fullpage}

\makeatletter \@addtoreset{equation}{section}
\makeatother

\begin{document}
\begin{titlepage}
 \thispagestyle{empty}
% \begin{flushright}
 %    \hfill{CERN-PH-TH/2009-150}\\
 %\end{flushright}

 \vspace{250pt}

 \begin{center}
     { \LARGE{\bf      {Black Holes and Attractors in Supergravity}}}

     \vspace{70pt}

     {\large {\bf Anna Ceresole$^{a,b}$ and Sergio Ferrara$^{b,c}$ }}

     \vspace{30pt}

   { \small  {\it ${}^a$ INFN, Sezione di Torino,
       Via Pietro Giuria 1, 10125 Torino, Italy\\E-mail: ceresole@to.infn.it} 

     \vspace{10pt}

  {\it ${}^b$ Physics Department, Theory Unit, CERN,\\
     CH -1211, Geneva 23, Switzerland\\E-mail: sergio.ferrara@cern.ch}}

     \vspace{10pt}

  {\small  {\it ${}^c$ INFN - Laboratori Nazionali di Frascati,\\ 
     Via Enrico Fermi 40, I-00044 Frascati, Italy}  }

     \vspace{15pt}

     \vspace{100pt}

     {ABSTRACT}
 \end{center}

 \vspace{10pt}
\noindent
We discuss some of the basic features of extremal black holes in four-dimensional extended supergravities. Firstly, all regular solutions display an attractor behavior for the scalar field evolution towards the black hole horizon. Secondly, they can be obtained by solving first order flow equations even when they are not supersymmetric, provided one identifies a suitable superpotential W which also gives the black hole entropy at the horizon and its ADM mass at spatial infinity. 
% U-Duality of the underlying supergravity constrains W and allows for a classification of the % solutions in terms of orbits of the electric-magnetic charge vector.
 We focus on N=8 supergravity and we review the basic role played by  U-duality of the underlying supergravity in determining the attractors, their entropies, their masses and  in classifying both regular and singular extremal black holes.
\vfill
\begin{center}{\sl
Contribution to the Proceedings of the Conference in Honor of Murray Gell-Mann's 80th Birthday,
Singapore, 24th-26th February 2010\\
Lecture delivered by Sergio Ferrara}
\end{center}

\end{titlepage}

\baselineskip 6 mm

%\section{Introduction} \label{Intro}

%\newpage\tableofcontents

\section{Extremal Black Holes}

In 1976, at the dawn of the N=1 supersymmetric theory of gravity in four dimensions \cite{Sugra}(now called ``N=1 supergravity''), when its N=2 extension had just appeared, Murray Gell-Mann was the first to remark during a seminar at Caltech that, if higher N supergravity would have indeed existed, then there would be a bound such that 
$N_{\text{max}}$=8 would be the end of the story.
Today, after 34 years, we are still struggling to understand this beautiful maximally extended theory\cite{CJ}, its connection with superstring and M-theory\cite{CJS}, its hypothetical perturbative finiteness and its non perturbative completion.  This contribution focusses on the black holes that arise in N=8 supergravity in four dimensions\cite{Ferrara:2006em}, which have been recently claimed to play a possible key role in relation to string theory and to the issue of perturbative finiteness of N=8 supergravity\cite{Dixon}.

Black holes, one of the most interesting outcome of General Relativity, are the typical probes of the quantum regime of any fundamental theory of gravity and  as such, they are naturally investigated within the framework of superstring and M-theory. As a first approximation, they can be proficiently studied as classical solutions of the underlying extended supergravities, which arise upon compactification in the effective field theory limit. Once the background geometry of the d-dimensional spacetime  and the number N of supersymmetry charges have been selected, all the important features of a given solution are encoded into the electric magnetic-duality group G acting on the vector fields $A^\Lambda$, and in the geometric properties of the moduli space G/H parametrized by the scalar fields $\phi^i$\cite{GZreview} . When the electric and magnetic charges are quantized, the group G becomes the U-duality group which is known to dictate the string dynamics in various dimensions\cite{Cvetic:1996zq,Witten}. 

The thermodynamical properties of black holes can be obtained  from quantum mechanical attributes that are their (ADM) mass, charge, spin and scalar charges (see for instance \cite{Ferrara:2008hwa,SUSY,FGK,reviewSF}). 
Unlike Schwarzschild black holes, charged (Reissner-Nordstrom) and/or spin (Kerr-Newman) black holes can be {\sl extremal}, {\sl i.e} with vanishing temperature for non-zero entropy, in which case their event  horizon and Cauchy horizons coincide. In formulae, the extremality parameter is given by
\begin{equation}
c=2ST=\frac12(r_+-r_-)\rightarrow 0\, ,
\end{equation}
where $c$ measures the surface gravity and $S=log\ {\cal N}$ is the black hole entropy which counts the number ${\cal N}$ of microstates. In supergravity, it is given by the Bekenstein-Hawking area formula
\begin{equation}
S_{BH}=\frac12 A_H=\pi R^2_H\,  ,
\end{equation}
where $R_H$ is the effective radius of a sphere encircling the horizon.  For extremal charged black holes, $R_H$ must respect the symmetries of the theory and in particular it must depend only on the electric and magnetic charges and not on the scalar field values  \cite{GZreview}.  Therefore, also the entropy will depend only on the charges and it will take particular expressions depending on the duality symmetries of the given model.

The lack of  dependence on scalar fields  of the entropy can be viewed as a sort of ``no hair" theorem \cite{Kallosh:1992ii} and reflects the fact that, under certain conditions, extremal black holes in N-extended supergravities enjoy a remarkable property: the scalar field trajectories $\phi^i (\tau)$, in terms of the radial evolution parameter $\tau=-1/\rho$  from asymptotic infinity ($\tau\rightarrow 0$) to the horizon ( $\tau\rightarrow -\infty$), behave as dynamical systems, {\sl i. e.} independently on their initial values $\phi^i_0=\phi^i(\tau)|_{\tau\rightarrow 0}$. The scalars evolve to a common value at the horizon, $\phi^i_H=\phi^i_H (Q)$, which they reach with zero velocity ($\dot\phi^i\rightarrow 0$), and where they entirely depend on the electric-magnetic charge vector $Q$ of the asymptotic configuration. This attractive feature is called {\sl Attractor Mechanism} \cite{SUSY,FGK}, and the attractor fixed points can be obtained as extrema of a suitable effective potential. 
Another important feature of extremal black holes is that their horizon geometry of spacetime is universal, and in four dimensions it is given by the  $AdS_2\times S_2$ Bertotti-Robinson metric. 
This is a particular case of the geometry of black p-branes in D-dimensions and it is an instance of the AdS/CFT correspondence relating the gauge theory on the boundary to bulk supergravity\cite{Aharony:1999ti}.
Thus, extremal black holes behave as solitons interpolating between maximally symmetric geometries of (super)spacetime: Minkowski for $\tau\rightarrow 0$ and the conformally flat metric for $\tau\rightarrow -\infty$ \cite{Duff:1994an,GibbTown}. 

It is remarkable that the common radius  of $AdS_2\times S_2$ and therefore the entropy, can be actually computed using the electric-magnetic duality of the underlying supergravity theory.

For Reissner-Nordstrom black holes,  with electric charge $q$ and magnetic charge $p$, the entropy is $S\sim p^2+q^2$ due to the $U(1)$ symmetry that rotates the p and q charges into each other. For  more complicated objects where scalar fields are present, such as the Axion-Dilaton black hole  having two $U(1)$ gauge fields and charges $(q_1,p_1)$ and $(q_2,p_2)$, the entropy becomes $S\sim |p_1q_2-p_2 q_1|$ and it is invariant under $SL(2)\times SO(2)$ \cite{Kallosh:1992wa}. The appearance of a non compact symmetry group is a quite general signature of the presence of scalar fields.
In $N=4$ supergravity, one has $S\sim |p^2q^2-(p.q)^2|^{1/2}$, with $SL(2)\times SO(6,n)$ symmetry\cite{Cvetic2}.
In the maximal $N=8$ case, the dyonic charge vector $Q^a$ transforms in the fundamental ${\bf 56}$ representation of $E_{7(7)}$  (since there are 28 vector fields yielding  28 electric and 28 magnetic charges) while the scalar fields span the 70-dimensional  scalar manifold $G/H=E_{7(7)}/SU(8)$. It turns out that the entropy for regular black holes is proportional to $ \sqrt{|I_4|}$ where $I_4$ is the quartic invariant of  the $\bf 56$ representation of  $E_{7(7)}$, $I_4=T_{abcd}Q^aQ^bQ^cQ^d$. When $I_4=0$, the black hole is singular, with vanishing horizon area, and there can be $1/8$, $1/4$ or $1/2$ supersymmetry preserved \cite{FG,Kallosh:1996uy,Arcioni:1998mn}.

In the last few years it has become clear that the scalar field dynamics for the extremal black holes can be entirely  encoded into a real ``superpotential'' function $W(\phi^i,Q)$ for which $\phi^i_H$ is a critical point in the moduli space of the theory: $\partial W/\partial \phi^i|_{\phi^i=\phi^i_H}=0$.

For BPS (supersymmetric) configurations, $W=|z_h|=\rho_h$ where $\rho_h$ is the modulus of the highest among  the skew eigenvalues  of the central charge matrix $Z_{AB}=-Z_{BA}$ of the supersymmetry algebra,
\begin{eqnarray}
\{{\cal Q}_{\alpha A},{\cal Q}_{\beta B}\}&=&\epsilon_{\alpha\beta}Z_{AB}(\phi^i_0,Q)\\
\{{\cal Q}_{\alpha A},{\cal Q}_{\dot\beta}^ B\}&=&\sigma^\mu_{\alpha\dot\beta}P_\mu \delta_A^B\, .
\end{eqnarray}
Depending on how many of these skew eigenvalues are coincident, one has various degrees of preserved supersymmetry. For the N=8 theory, one can have four different cases ranging from $1/8$, to $1/4$, $1/2$ and zero preserved supersymmetries.
In the original setup of N=2 extremal black holes in D=4, $Z_{AB}=\epsilon_{AB}Z$ and such a superpotential function for supersymmetric configurations was to be identified with the modulus of the N=2 central charge $Z$, appearing in the ordinary BPS equations \cite{SUSY,FGK}.
Remarkably, such a function W can be shown to exist also for non supersymmetric configurations, in which case it is called the ``fake superpotential'' \cite{Ceresole:2007wx} because of the similarity with the set up of ``fake supergravities" \cite{fake}.
When the attractors are regular, the W function has a minimum for $\phi^i=\phi^i_H$, and its horizon value gives the entropy of the configuration 
\begin{equation}
S=\frac14 A_H=\pi W_H^2(Q)=\pi W^2_{\mathrm{crit}}(\phi_h^i(Q),Q)
\end{equation}
according to the Bekenstein-Hawking formula. However, the if W has a runaway behavior in moduli space, $\phi_H\to\infty\, W\to 0$ (which is not acceptable in D=4), the corresponding black hole solutions are singular. Then the scalar fields are never stabilized within the boundaries of moduli space, there are no attractors and the entropy of the extremal configuration vanishes.

In order to describe a static, spherically symmetric extremal black hole background in the extremal case, $c=2ST=0$,  the metric ansatz reads \cite{FGK} 
\begin{equation}
ds^2=-e^{2U}dt^2+e^{-2U} \frac{d\tau^2}{\tau^4}+\frac1{\tau^2} (d\theta^2+\sin{\theta}^2d\phi^2)\ .
\end{equation}
with the field strength $F_{\mu\nu}^\Lambda$  for $n_v$ vectors ($\Lambda=1,\ldots,n_V$) and its dual $G_{\Lambda\mu\nu}=\frac{\delta{\cal L}}{\delta F_{\mu\nu}^\Lambda}$ given by
\begin{eqnarray}
F&=&e^{2U}C{\cal M}(\phi^i)Qdt\wedge d\tau+Q\sin{\theta}d\theta\wedge d\phi\, \\
F&=&
\left(
\begin{array}{c}
  F^\Lambda_{\mu\nu}  \\
  G_{\Lambda\mu\nu}
\end{array}
\right)
\frac{dx^\mu dx^\nu}{2}\, .
\end{eqnarray}
 Electric and magnetic charges are defined by
\begin{equation}
q_\Lambda=\frac1{4\pi}\int_{s_2} G_\Lambda\, ,\qquad p^\Lambda=\frac1{4\pi}\int_{S_2} F_\Lambda\, .
\end{equation}
\noindent
 ${\cal M} (\phi^i)$ is  a $2n_v\times 2n_v$ real symmetric $Sp(2n_v,R)$ matrix, satisfying ${\cal M} C{\cal M}=C$ ,
\begin{equation}
{\cal M}(\phi^i)=
\left(
\begin{array}{cc}
  I+R I^{-1}R  &-RI^{-1}   \\
-I^{-1}R  &   I^{-1}   
\end{array}\, .
\right)
\end{equation}
where $I=Im\,{\cal N}_{\Lambda\Sigma},R= Re\,{\cal N}_{\Lambda\Sigma}$, the vector kinetic matrix ${\cal N}_{\Lambda\Sigma}$ depends on the scalar fields and enters the 4D lagrangian
\begin{equation}
{\cal L}=-\frac{R}2+\frac12g_{ij}(\phi)\partial_\mu \phi^i\partial^\mu \phi^j+ I_{\Lambda\Sigma}F^{\Lambda}\wedge ^*F^{\Sigma}+R_{\Lambda\Sigma}F^{\Lambda}\wedge F^{\Sigma}\, .
\end{equation}

The black hole effective potential \cite{SUSY} is given by
\begin{equation}
V_{BH} = -\frac12 Q^T {\cal M} Q=\frac12 Z_{AB}{\overline Z}^{AB} =\sum_i \rho_i^2\, ,\label{eff}
\end{equation}
where A,B are SU(8) indices \cite{reviewSF}. In the last expression, $\rho_i$ are the moduli of the skew eigenvalues  $z_i$ of the central charge matrix $Z_{AB}$. It arises upon reducing the general 4D (or higher D) lagrangian to  the one-dimensional almost geodesic action describing the radial evolution of the $n+1$ scalar fields $(U(\tau),\phi^i(\tau))$:
\begin{equation}
S=\int{{\cal L}d\tau}=\int (\dot{U}+g_{i\bar\jmath}\dot{\phi}^i\dot{\phi}^{\bar\jmath}+e^{2U}V_{BH}(\phi(\tau),p,q)d\tau.
\end{equation}
In order to have the same equations of motion of the original theory, the action must be complemented with the Hamiltonian constraint (in the extremal case) \cite{FGK}
\begin{equation}
\dot{U}^2+g_{ij}\dot{\phi}^i\dot{\phi}^{j}-e^{2U}V_{BH}(\phi(\tau),p,q)=0\, .
\end{equation}
The black hole effective potential can be written in terms of the superpotential $W(\phi)$ as 
\begin{equation}
V_{BH} = W^2+2g^{i{j}}\partial_i W \partial_{j} W\, \label{VW}\, .
\end{equation}
This formula can be viewed as a differential equation defining W for a given black hole effective potential $V_{BH}$, and it can lead to multiple choices: only one of those will corresponds to BPS solutions, while a different one will be associated to non BPS ones. In both cases, W allows to rewrite the ordinary second order supergravity equations of motion
\begin{eqnarray}
\frac{d^2U}{d\tau^2}&=&e^{2U} V_{BH}\\
\frac{d^2\phi^i}{d\tau^2}&=&g^{i\bar\jmath}\frac{\partial V_{BH}}{\partial\phi_{\bar\jmath}}e^{2U}\, ,
\end{eqnarray}
as first order flow equations, defining the radial evolution of the scalar fields $\phi^i$ and the warp factor $U$ from asymptotic infinity towards the black hole horizon \cite{Ceresole:2007wx} :
\begin{equation}
U^\prime =-e^UW\, , \qquad \qquad \phi^{\prime i}=-2e^Ug^{i{j}}\partial_{j} W\, . 
\end{equation}
\noindent
The important point is that, at the prize of finding a suitable fake superpotential W, one only has to deal with these first order flow equations  even for non supersymmetric solutions, where one does not have Killing spinor equations \cite{Ceresole:2007wx,Andrianopoli:2007gt}.

Beside the horizon entropy $S_{BH}=\pi W^2_H$ and the first order flows, the value at radial infinity of the superpotential $W$ also encodes other basic property of the extremal black hole, which are its  $ADM$ mass, given by
\begin{equation}
M_{ADM}(\phi_0^i,Q)=\dot U(\tau=0)=W(\phi_0,Q)
\end{equation}
and the scalar charges at infinity
\begin{equation}
\Sigma^i=\dot\phi^i(\tau=0)=2g^{ij}(\phi_0)\frac{\partial W}{\partial \phi^i} (\phi_0,Q)
\end{equation}
A finite horizon area demands that $\dot\phi^i(\tau=-\infty )=0$ and thus $\frac{\partial W}{\partial \phi^i}|_{\phi^i_H}=0$ and
\begin{equation}
\lim_{\tau\to-\infty} e^{-2U}=R^2_H\tau^2\, ,R^2_H=W^2_H(\phi_H,Q)
\end{equation}
and $R^2_H(Q)=|I_4(Q)|^{1/2}$, so that indeed the effective radius is given in terms of the Cartan quartic invariant of the $56$ of $E_{7(7)}$.

By using the asymptotic behaviour of the warp factor at spacial infinity $\tau \to 0$, one has $U\to -M_{ADM}\tau$ and at the horizon $\tau\to\infty$
\begin{equation}
e^{-2U}\to\frac1{4\pi} A_H\tau^2
\end{equation}
Thus, one gets model independent expressions for ADM mass, scalar charges and  entropy as a function of W
\begin{eqnarray}
M^2_{ADM} &=&W^2_{\tau\to 0}=W^2_{\infty}\\
\Sigma^i &=& -2g^{ij}\frac{\partial W}{\partial \phi^j}|_{\tau\to 0}=-2g^{ij}\frac{\partial W}{\partial \phi^j}|_{\infty}\\
S&=&\frac14A_H = \pi W^2|_{\tau\to -\infty}=\pi W^2|_{H}
\end{eqnarray} 

The horizon value of the scalar fields $\phi_H^i$ is a critical point for $W$ in the moduli space of the theory, $\partial W/\partial\phi^r|_{\phi^i=\phi^i_H}=0$.
It follows that for $\tau\to 0$ (radial infinity)
\begin{equation}
M^2_{ADM}=V_{BH} (\phi_\infty, p, q)-\frac12 g_{ij} \Sigma^i\Sigma^{j}
\end{equation}
while for $\tau\to-\infty$ (horizon)
\begin{equation}
S_{BH}=\frac14 A_H=\pi V_{BH}(\phi_H,p,q)=\pi V_{BH}|_{\phi_{H}}
\end{equation}
showing {\sl attractor behaviour} \cite{SUSY}.
For BPS states
\begin{equation}
S_{BH}=\pi |z_{h}|^2_{H}\, \quad\quad M_{ADM}^2=|z_{h}|^2_{\infty}\, ,
\end{equation}
where $z_h$ is the skew eigenvalue of the central charge $Z_{AB}$ with the highest modulus.
Quite generally, one can prove the validity of the BPS bound along the flow
\begin{equation}
M_{ADM}\geq|z_i|\, .
\end{equation}
At the horizon
\begin{equation}
\dot\phi^i|_{\tau\to-\infty}=0\rightarrow \partial_iW_H=0\quad (W_H\leftrightarrow W_{crit})\, .
\end{equation}

For $N\geq 2$, the fake superpotential  for the non-BPS branch\cite{Dabholkar:2006tb,larsen} has been computed for wide classes of models  \cite{Ceresole:2007wx,Cardoso5d,Andrianopoli:2007gt,Bellucci:2008sv,Ceresole:2009iy,universality, Bossard:2009we, varinonBPS,Andrianopoli:2009je}, based on symmetric geometries of moduli spaces, using as a tool the U-duality symmetry of the underlying supergravity. A universal procedure  for its construction  in $N=2$ special geometries has been established \cite{Ceresole:2009iy,universality}, which generalizes  the results obtained for the $stu$ model\cite{stu}.  This universal procedure can be also applied to the $N=8$ theory using the fact that  the $stu$ model is both a subsector of $N=8$ and a model in $N=2$ where we know how to describe $W$ in terms of invariants. The results of \cite{universality} agree with the outcome of studies of black hole evolution with the method of time-like reduction to 3 dimensions in \cite{Bossard:2009we}. However, in both contexts, the expression for W is only given implicitly, as solution of a sixth order polynomial having for coefficients some $SU(8)$ invariant functions composed out of the $N=8$ central charge.

\section{Attractors and Duality orbits in N=8 supergravity}
 
\noindent

As already noticed above, U-duality of the underlying  extended supergravity dictates many important features \cite{Cvetic:1996zq} .
Another important point is that the absolute  Cartan invariant $I_4$ of the $56$ dimensional representation of $E_{7(7)}$ and its derivatives allow to specify the supersymmetric features of a given solution and to classify the U-duality (continuous) orbits for the charge vector Q \cite{Ferrara:1997ci, orbits,orbits2}.  

The area of the horizon for both regular 1/8-BPS and non-BPS attractors, is proportional to the square of the absolute Cartan quartic invariant \cite{Kallosh:1996uy}
\begin{equation}
I_4=Tr(Z{\overline Z})^2-\frac14(Tr\, Z{\overline Z})^2+4(Pf\,  Z+Pf\, {\overline Z})\, .
\end{equation} 
This expression, involving the traces of the matrix $Z_{AB}{\bar Z}^{BC}$ and of its square as well as the real part of the Pfaffian of  $Z_{AB}(\phi,Q)$,
\begin{equation}
Pf Z=\frac1{2^4 4!}\epsilon^{abcdefgh}Z_{AB}Z_{CD}Z_{EF}Z_{GH}
\end{equation}
actually depends only on the electric and magnetic charges $Q=(p,q)$ and not on the scalar fields.
Moreover, for fixed values of  $I_4$ in d=4 and of an analogous cubic invariant $I_3$  in d=5, charge vectors Q for supergravities on symmetric spaces describe related  orbits and attractors \cite{Ceresole:2009id, Ceresole:2009jc}.

By an $SU(8)$ transformation, one can reach the canonical bases where the antisymmetric central charge matrix is (skew)diagonalized  and takes the form
\begin{equation}
Z_{AB}=
\left(
\begin{array}{cccc}
  \rho_0 &   & &   \\
  &  \rho_1 & &   \\
  &   & \rho_2&\\ & & & \rho_3 
\end{array}
\right)\otimes \left(
\begin{array}{cc}
  0&  1   \\
  -1& 0 
\end{array}
\right)e^{i\varphi/4}
\end{equation}
in terms of its skew eigenvalues $z_i=\rho_i e^{i\varphi/4}$, $i=0,1,2,3$ which amount to 5 independent parameters.
The attractor equation becomes
\begin{equation}
\partial_iV=0 \rightarrow z_i z_j+z^*_kz_l^*=0\, , (i\neq j\neq k \neq l)
\end{equation}
which yields two solutions
\begin{itemize}
\item[a)] $z_h\neq 0$, $z_{i\neq h}=0$
\item[b)] $|z_i|=\rho$ , $ arg Pf Z =\pi $
\end{itemize}
Considering the quartic invariant in this basis,
\begin{eqnarray}
I_4&=&[(\rho_0+\rho_1)^2-(\rho_2+\rho_3)^2][(\rho_0-\rho_1)^2\\
&-&(\rho_2-\rho_3)^2]  +8\rho_0\rho_1\rho_2\rho_3(\cos{\phi}-1)\\
&&\rho_0\geq\rho_1\geq\rho_2\geq\rho_3
\end{eqnarray}
one finds that there are two disjoint regular orbits corresponding to the a) and b) attractor solutions:

\noindent
i) 1/8 BPS: with $I_4>0$, for  $\rho_0\neq 0$ and $\rho_1=\rho_2=\rho_3=0$. The duality orbit is $E_{7(7)}/E_{6(2)}$.

\noindent
ii) non BPS: with  $I_4<0$, for  $|z_i|=\rho$  and $ \mathrm{arg} Pf\,  Z =\pi $. The  duality orbit is $E_{7(7)}/E_{6(6)}$.

\noindent
There are three additional ``small" orbits, preserving respectively 1/8, 1/4 and 1/2 supersymmetry, which are singular, having $I_4=0$ and thus zero horizon area.

The entropy of the two regular branches is given by
\begin{eqnarray}
S_{\text{1/8BPS}} &=&\pi V_{BH}|_{H}=\pi \rho_0^2|_{H}\, ,\\
S_{\text{nonBPS}} &=& \pi V_{BH}|_{H}= \pi (2\rho_0)^2|_H\, ,
\end{eqnarray}
where $\rho_0|_H$ is the fixed point value of the field dependent modulus  of the highest eigenvalue $\rho_0$.
\noindent
A similar classification has recently been achieved also in the $N=2$  case, where the superpotential W was computed for large and small orbits \cite{new}. The new feature in that case is that singular orbits for N=2 black holes can also be non-BPS and that W can always be determined by simple radicals.

An interesting outcome of the study of black holes in N=8 four dimensional supergravity is related to some recent results \cite{BFK}, which  point to a possible key role played by  black holes with singular geometry in relation to String Theory. N = 8 supergravity in D = 4 can be obtained by dimensional reduction of D=11, N=1 supergravity (M-theory) on a 7-torus $T^7$, or compacifying Type-II string theory on $T^6$. In \cite{GOS} it has been recently observed that  in the process of compactification, one is left not only with the 256 massless states of the N=8, D=4 supergravity, but also with an infinite tower of stringy  elementary  states with arbitrarily small mass, making the decoupling impossible. 
On the other hand, since the degeneracies of  BPS states in string and M-theory are counted by U-duality invariant formulas, it is interesting to inquire whether one
can  consistently decouple these extra massless states without violating the U-duality invariance of the degeneracy formula.  If one can't decouple the extra states without breaking U-duality, then this would suggest that one may be able to disprove the conjecture of UV finiteness of the perturbative  N=8 supergravity theory. From the 4D point of view, these additional massless states appear to correspond to classical black hole solutions carrying charges which lead to light-like orbits with $I_4=0$ \cite{BFK}, and  that $N=8$ black holes with vanishing area of the horizon can be interpreted as dynamically reduced black branes with regular geometry. For example, 1/8 BPS black holes with $I_4=0$ correspond to reduced (wrapped) black holes (black strings) at D=5 (with three charges $q_1,q_2,q_3$, $J_{MAX}=7/2$). Conversely, 1/4 BPS black holes with $\partial I_4=0$ correspond to reduced and wrapped black strings at $D=6$ (two charges $q_1$ and $q_2$, $J_{MAX}=3$. Finally, 1/2 BPS black holes with $\partial ^2 I_4=0$ correspond to KK states i.e. massless particles in higher dimensions $J_{MAX}=2$. If only regular black holes, with $AdS_2\times S_2$  horizon geometries are to be retained in a consistent theory of gravity, one may not need to open up  extra dimensions\cite{GOS} : weather this feature could hint to a possible finite theory of D=4, $N=8$ supergravity \cite{Dixon}  without the need of string theory, is a question that is presently under close investigation.

 \section{Moduli spaces of attractors and flat directions}
 It is a property of N=2 supergravity that the geometry of black holes, at least in the Einsteinian approximation, does not depend on the hypermultiplet moduli space\cite{SUSY}. This implies that attractive solutions exist for arbitrary v.e.v's of the hypermultiplet scalars. This phenomenon generalizes to the non-BPS N=2  regular black holes as well as to small black holes, for which there is no attractor behaviour \cite{Ferrara:2007tu,Andrianopoli:2010bj, new}. In fact, it has been realized that, in terms of the fake superpotential W (or, for BPS solutions, in terms of the $Z$ central charge), such ``flat directions'' are flat directions for W along the complete flow and therefore the ADM mass as well. For $N>2$, flat directions exist even for BPS regular attractive (large) black holes and for singular solutions. In this latter case, the moduli space of flat directions is usually bigger, and in fact for the doubly critical orbits the W function just depends on one real modulus having the simple interpretation as the radius dependence of the Kaluza-Klein mass \cite{Cerchiai:2010xv,Borsten:2010aa}. Let us confine again the discussion to the N=8 case. The five charge orbits are as follows \cite{FG}
 \begin{equation}
\begin{array}{|l|l|l|l|}
\hline
  \text{orbit}&\text{susy}& \text{coset}  &\text{moduli space}   \\
  \hline
 I_4>0 & 1/8 \text{BPS}\qquad &\frac{ E_{7(7)} }{ E_{6(2)} }   & \frac{ E_{6(2)} }{ SU(6)\times SU(2)}   \\
I_4<0 &\text{nonBPS}\qquad  &\frac{ E_{7(7)}}{E_{6(6)} }   &   \frac{E_{6(6)} }{USp(8)}  \\
I_4=0,\partial I_4\neq 0&1/8\text{BPS}\qquad&\frac{E_{7(7)} }{F_{4(4)}\odot T_{26} }&
\frac{F_{4(4)}}{USp(6)\times USp(2)}\odot T_{26}\\
I_4=0,\partial I_4= 0&1/4\text{BPS}&\frac{E_{7(7)}}{O(6,5)\odot (T_{32}+T_1)}&\frac{O(6,5)}{O(6)\times O(5)}\odot ( T_{32}+T_1)\\
I_4=0,\partial I_4= 0,\partial^2 I_4=0\quad &1/2\text{BPS}\quad&\frac{E_{7(7)}}{E_{6(6)}\odot T_{27}}\qquad\qquad&\frac{E_6(6)}{USp(8)}\odot T_{27}\\\hline
\end{array}
 \end{equation}
 The last column refers to the moduli space of flat directions and is given by the stabilizer of the coset orbit divided by its maximal compact subgroup.
One observes that small lightlike black holes, having only the condition $I_4=0$, apart from $T_{26}$ translations, have the same moduli space as the 1/8 BPS large black hole in D=5 \cite{Ferrara:2007tu}. 

Remarkably, the 1/2 BPS black holes, apart from $T_{27}$ additional translations, have the same moduli space as large non-BPS black holes and both are the 5D N=8 supergravity moduli space. This is due to the fact that both these configurations can be obtained without genuine 5D black hole charges  but  turning on the NUT charge and angular momentum , without breaking the $E_{6(6)}$ symmetry of the 5D theory\cite{Cardoso5d,Ceresole:2009id,Ceresole:2009jc}.

\section{ Conclusions and Other recent developments}
Since the attractor mechanism was observed\cite{SUSY}, there has been a lively activity around the study of black holes in connection with string theory  and M-theory.  To begin with, an intriguing connection was  conjectured  between the central charge and the topological supergravity partition function at the attractor point \cite{OSV}. A further important development has been the computation of the entropy through microstate counting, reproducing the Bekenstein-Hawking formula in the regime of large charges \cite{SV}. 

Then, quantum corrections to the entropy formulae and to the attractor equations were computed in superstrings and supergravity with N=2 supersymmetry \cite{Maldacena:1997de,DeWit,Sen,Maloney}.
Microstate counting for black holes was considered in many papers, including the case of small black holes for which the area formula does not apply, even in the classical regime \cite{Dabholkar,Shih:2005qf,Dabholkar:2005dt}.
Several aspects concerning multicenter black holes, black hole deconstructions, split attractor flows and walls of marginal stability have been thoroughly investigated \cite{Denef:2007vg, Denef:2007yt,Sen:2008ht,Gaiotto:2009hg}. 

The connection between 5D and 4D  has been explored under different perspectives, as for instance in \cite{Gaiotto:2005xt,Ceresole:2009id} among many other contributions.
Three dimensional time reduction has also been shown to be a powerful tool for black hole classification, duality orbits and possible quantization of the geodesic flows \cite{Gun,geo,Bossard:2009at}.

As a final note, we would like to mention another more speculative aspect of black hole physics, namely the mathematical connection between black hole physics and Quantum Information Theory, in particular the relation between black hole entropy and qubits multipartite entanglement \cite{DuffQIT}.

\section*{Acknowledgments}
This work is supported in part by the ERC Advanced Grant no. 226455, \textit{``Supersymmetry, Quantum Gravity and Gauge Fields''} (\textit{SUPERFIELDS}), by MIUR-PRIN contract 20075ATT78 and by DOE Grant DE-FG03-91ER40662.


\begin{thebibliography}{[1]}

\bibitem{Sugra} S. Ferrara, D. Z. Freedman and P. Van Nieuwenhuizen, {\it Progress Toward a Theory of Supergravity}, Phys. Rev. {\bf D13} (1976) 3214;
S. Deser and B. Zumino, {\it Consistent Supergravity}, Phys. Lett. {\bf 62B} (1976) 335.

\bibitem{CJ} 
 %\cite{Cremmer:1979up}
%\bibitem{Cremmer:1979up}
  E.~Cremmer and B.~Julia,
  {\it The SO(8) Supergravity},
  Nucl.\ Phys.\  B {\bf 159} (1979) 141.
 
 \bibitem{CJS} 
 %\bibitem{Cremmer:1978km}
  E.~Cremmer, B.~Julia and J.~Scherk,
  {\it Supergravity theory in 11 dimensions}
  Phys.\ Lett.\  B {\bf 76} (1978) 409.
  %%CITATION = PHLTA,B76,409;%%
 
 %\cite{Ferrara:2006em}
\bibitem{Ferrara:2006em}
  S.~Ferrara and R.~Kallosh,
  {\it On N = 8 attractors},
  Phys.\ Rev.\  D {\bf 73}, 125005 (2006)
  [arXiv:hep-th/0603247].
  %%CITATION = PHRVA,D73,125005;%%

\bibitem{Dixon}
%\cite{Bern:2006kd}
%\bibitem{Bern:2006kd}
  Z.~Bern, L.~J.~Dixon and R.~Roiban,
  {\it Is N = 8 Supergravity Ultraviolet Finite?},
  Phys.\ Lett.\  B {\bf 644}, 265 (2007)
  [arXiv:hep-th/0611086];
  %%CITATION = PHLTA,B644,265;%%
%\cite{Bern:2009kd}
%\bibitem{Bern:2009kd}
  Z.~Bern, J.~J.~Carrasco, L.~J.~Dixon, H.~Johansson and R.~Roiban,
  {\sl The Ultraviolet Behavior of N=8 Supergravity at Four Loops},
  Phys.\ Rev.\ Lett.\  {\bf 103}, 081301 (2009)
  [arXiv:0905.2326 [hep-th]].
%\cite{Andrianopoli:2010bj}

\bibitem{GZreview}
%\cite{Aschieri:2008ns}
%\bibitem{Aschieri:2008ns}
%\bibitem{Gaillard:1981rj}
  M.~K.~Gaillard and B.~Zumino,
  {\it Duality Rotations For Interacting Fields},
  Nucl.\ Phys.\  B {\bf 193} (1981) 221;
  P.~Aschieri, S.~Ferrara and B.~Zumino,
  {\it Duality Rotations in Nonlinear Electrodynamics and in Extended
  Supergravity},
  Riv.\ Nuovo Cim.\  {\bf 31}, 625 (2009)
  [arXiv:0807.4039 [hep-th]].

\bibitem{Cvetic:1996zq} C. M. Hull and P. K. Townsend, Nucl.\  Phys.\  B{\bf  438} 109 (1995); M.~Cvetic and C.~M.~Hull, 
{\it Black holes and U-duality}  
Nucl.\ Phys.\  B {\bf 480} (1996) 296   [arXiv:hep-th/9606193].

\bibitem{Witten}%\cite{Witten:1995ex}
%\bibitem{Witten:1995ex}
  E.~Witten,
  {\it String theory dynamics in various dimensions},
  Nucl.\ Phys.\  B {\bf 443} (1995) 85
  [arXiv:hep-th/9503124].
  %%CITATION = NUPHA,B443,85;%%

\bibitem{Ferrara:2008hwa}
  S.~Ferrara, K.~Hayakawa and A.~Marrani,
  {\it Lectures on Attractors and Black Holes},
  Fortsch.\ Phys.\  {\bf 56}, 993 (2008)
  [arXiv:0805.2498 [hep-th]].
 
\bibitem{SUSY}  S.~Ferrara, R.~Kallosh and A.~Strominger,
 {\it N=2 extremal black holes}, 
Phys.\ Rev.\ D {\bf 52} (1995) 5412; 
A. Strominger,
 {\it Macroscopic entropy of N=2 extremal black holes},
 Phys. Lett. {\bf B383}, 39 (1996);
S. Ferrara and R. Kallosh, 
{\it Supersymmetry and attractors}, 
Phys. Rev. D {\bf 54}, 1514 (1996);
S. Ferrara and R. Kallosh, 
{\it Universality of supersymmetric attractors}, 
Phys. Rev. D {\bf 54}, 1525 (1996).

\bibitem{FGK}
S. Ferrara, G. W. Gibbons and R. Kallosh, 
{\it Black holes and critical points in moduli space}, 
Nucl. Phys. B {\bf 500}, 75 (1997).
 
\bibitem{reviewSF}L. Andrianopoli, R. D'Auria, S. Ferrara and M. Trigiante, {\it Extremal Black Holes in Supergravity}, Lect.Notes Phys.737:661 (2008), [arXiv:hep-th/0611345]

%\cite{Kallosh:1992ii}
\bibitem{Kallosh:1992ii}
  R.~Kallosh, A.~D.~Linde, T.~Ortin, A.~W.~Peet and A.~Van Proeyen,
  {\it Supersymmetry as a cosmic censor},
  Phys.\ Rev.\  D {\bf 46}, 5278 (1992)
  [arXiv:hep-th/9205027].
  %%CITATION = PHRVA,D46,5278;%%

\bibitem{Aharony:1999ti}
  O.~Aharony, S.~S.~Gubser, J.~M.~Maldacena, H.~Ooguri and Y.~Oz,
  {\it Large N field theories, string theory and gravity},
  Phys.\ Rept.\  {\bf 323}, 183 (2000)
  [arXiv:hep-th/9905111].

\bibitem{Duff:1994an}
  M.~J.~Duff, R.~R.~Khuri and J.~X.~Lu,
  {\it String solitons},
  Phys.\ Rept.\  {\bf 259}, 213 (1995)
  [arXiv:hep-th/9412184].

\bibitem{GibbTown} G. Gibbons and P. K. Townsend, {\it Vacuum interpolation in supergravity via super p-brane},  Phys.Rev.Lett.{\bf 71}, (1993) 3754 . 

\bibitem{Kallosh:1996uy} R.~Kallosh and B.~Kol, 
{\it E(7) Symmetric Area of the Black Hole Horizon}, 
Phys.\ Rev.\  D {\bf 53}, 5344 (1996) [arXiv:hep-th/9602014].
%%CITATION = PHRVA,D53,5344;%%
%\cite{Arcioni:1998mn}
\bibitem{Arcioni:1998mn}
  G.~Arcioni, A.~Ceresole, F.~Cordaro, R.~D'Auria, P.~Fre, L.~Gualtieri and M.~Trigiante,
  {\it N = 8 BPS black holes with 1/2 or 1/4 supersymmetry and solvable Lie
  algebra decompositions},
  Nucl.\ Phys.\  B {\bf 542} (1999) 273
  [arXiv:hep-th/9807136].
  %%CITATION = NUPHA,B542,273;%%

\bibitem{Ceresole:2007wx}  A.~Ceresole and G.~Dall'Agata, 
{\it Flow Equations for Non-BPS Extremal Black Holes},
  JHEP {\bf 0703} (2007) 110     [arXiv:hep-th/0702088].

\bibitem{fake} 
%\bibitem{Freedman:2003ax}
  D.~Z.~Freedman, C.~Nunez, M.~Schnabl and K.~Skenderis,
  {\it Fake Supergravity and Domain Wall Stability},
  Phys.\ Rev.\  D {\bf 69} (2004) 104027
  [arXiv:hep-th/0312055].
  %%CITATION = PHRVA,D69,104027;%%
  %\cite{Celi:2004st}
%\bibitem{Celi:2004st}
  A.~Celi, A.~Ceresole, G.~Dall'Agata, A.~Van Proeyen and M.~Zagermann,
  {\it On the fakeness of fake supergravity},
  Phys.\ Rev.\  D {\bf 71}, 045009 (2005)
  [arXiv:hep-th/0410126].

%\cite{Dabholkar:2006tb}
\bibitem{Dabholkar:2006tb}
  A.~Dabholkar, A.~Sen and S.~P.~Trivedi,
  {\it Black hole microstates and attractor without supersymmetry},
  JHEP {\bf 0701}, 096 (2007)
  [arXiv:hep-th/0611143].
  %%CITATION = JHEPA,0701,096;%%
\bibitem{larsen}  E.~G.~Gimon, F.~Larsen and J.~Simon, 
{\it Black Holes in Supergravity: the non-BPS Branch},
 JHEP {\bf 0801} (2008) 040 [arXiv:hep-th/0710.4967].
%%CITATION = JHEPA,0801,040;%%

\bibitem{Andrianopoli:2007gt}  L.~Andrianopoli, R.~D'Auria, E.~Orazi and M.~Trigiante,
{\it First Order Description of Black Holes in Moduli Space},
JHEP {\bf 0711} (2007) 032	  [arXiv:hep-th/0706.0712].

\bibitem{Cardoso5d} G.~Lopes Cardoso, A.~Ceresole, G.~Dall'Agata, J.~M.~Oberreuter and J.~Perz,
 {\it First-order flow equations for extremal black holes in very special geometry},
JHEP {\bf 0710} (2007) 063 [arXiv:hep-th/0706.3373].

\bibitem{Bellucci:2008sv}  S.~Bellucci, S.~Ferrara, A.~Marrani and
A.~Yeranyan, {\it stu Black Holes Unveiled},  [arXiv:hep-th/0807.3503].
%%CITATION = ARXIV:0807.3503;%%

\bibitem{Ceresole:2009iy}
A.~Ceresole, G.~Dall'Agata, S.~Ferrara and A.~Yeranyan,
{\it  First order flows for N=2 extremal black holes and duality invariants,}
Nucl.Phys.B824:239-253,2010
[arXiv:0908.1110] .

\bibitem{universality}  A. Ceresole,  G. Dall'Agata, S. Ferrara and A. Yeranyan, {\it Universality of the superpotential for d=4 extremal black holes},  Nucl.\ Phys.  {\bf  B} (2010), in press   [arXiv:0910.2697].
 
\bibitem{Bossard:2009we} G.~Bossard, Y.~Michel and B.~Pioline, 
{\it Extremal black holes, nilpotent orbits and the true fake superpotential},
 [arXiv:0908.1742 ].
%\cite{Ceresole:2009id}

\bibitem{varinonBPS} S.~Ferrara, A.~Gnecchi and A.~Marrani, 
{\it d=4 Attractors, Effective Horizon Radius and Fake Supergravity,}
Phys.\ Rev.\  D {\bf 78} (2008) 065003 [arXiv:hep-th/0806.3196];
J.~Perz, P.~Smyth, T.~Van Riet and B.~Vercnocke, 
{\it First-order flow equations for extremal and non-extremal black holes,}
JHEP {\bf 0903} (2009) 150
[arXiv:hep-th/0810.1528];
K.~Goldstein and S.~Katmadas, {\it Almost BPS black holes,}
 JHEP {\bf 0905} (2009) 058 [arXiv:hep-th/0812.4183];
I.~Bena, G.~Dall'Agata, S.~Giusto, C.~Ruef and N.~P.~Warner, 
{\it Non-BPS Black Rings and Black Holes in Taub-NUT,}  
JHEP {\bf 0906} (2009) 015 [arXiv:hep-th/0902.4526];
P.~Galli and J.~Perz, 
{\it Non-supersymmetric extremal multicenter black holes with superpotentials,}
 [arXiv:0909.5185].

\bibitem{Ferrara:1997ci} S.~Ferrara and J.~M.~Maldacena, 
{\it Branes, central charges and U-duality invariant BPS conditions,} 
Class.\ Quant.\ Grav.\ {\bf 15}, 749 (1998) [arXiv:hep-th/9706097].

\bibitem{FG} S. Ferrara and M. Gunaydin, {\it Orbits of exceptional groups, duality and BPS states in string theory},  Int. J. Mod. Phys. A13 (1998) 2075; 

\bibitem{orbits}
S.~Bellucci, S.~Ferrara, M.~Gunaydin and A.~Marrani,
 {\it Charge orbits of symmetric special geometries and attractors,} 
Int.\ J.\ Mod.\ Phys.\  A {\bf 21} (2006) 5043 [arXiv:hep-th/0606209].

\bibitem{orbits2} L.~Andrianopoli, R.~D'Auria and S.~Ferrara, 
{\it U-duality and central charges in various dimensions revisited,} 
Int.\ J.\ Mod.\ Phys.\  A {\bf 13}, 431 (1998)
[arXiv:hep-th/9612105];\\
L.~Andrianopoli, R.~D'Auria and S.~Ferrara,
{\it Five dimensional U-duality, black-hole entropy and topological invariants,}  
Phys.\ Lett.\  B
{\bf 411}, 39 (1997)  [arXiv:hep-th/9705024].


\bibitem{Cvetic2}M. Cvetic and D. Youm, {\it Dyonic BPS saturated black holes of heterotic string on a six torus}, Phys. Rev. {\bf D 53} (1996) 584 [arXiv:hep-th/9507090].

%\cite{Kallosh:1992wa}
\bibitem{Kallosh:1992wa}
  R.~Kallosh, T.~Ortin and A.~W.~Peet,
  {\it Entropy and action of dilaton black holes},
  Phys.\ Rev.\  D {\bf 47} (1993) 5400
  [arXiv:hep-th/9211015].
  %%CITATION = PHRVA,D47,5400;%%

  \bibitem{democracy} %\cite{Townsend:1995gp}
%\bibitem{Townsend:1995gp}
  P.~K.~Townsend,
  {\it P-brane democracy},
  arXiv:hep-th/9507048.
  %%CITATION = HEP-TH/9507048;%%

%\bibitem{Strominger:1990pd}  A.~Strominger,
%``Special Geometry,''  
%Commun.\ Math.\ Phys.\ {\bf 133} (1990) 163.

%\bibitem{Strathdee} J. Strathdee, {``Extended Poincare Supersymmetry"}, Int. J. Mod. Phys.{\bf  %A2} (1) 273 (1987).

\bibitem{stu} M.~J.~Duff, J.~T.~Liu and J.~Rahmfeld, 
{\it Four-Dimensional String-String-String Triality,}
Nucl.\ Phys.\ B {\bf 459}, 125 (1996) [arXiv:hep-th/9508094]; K.~Behrndt, R.~Kallosh, J.~Rahmfeld, M.~Shmakova and W.~K.~Wong,
{\it STU black holes and string triality,} 
Phys.\ Rev.\ D {\bf 54} (1996) 6293 [arXiv:hep-th/9608059]. %%CITATION = PHRVA,D54,6293;%%

\bibitem{Andrianopoli:2009je}  L.~Andrianopoli, R.~D'Auria, E.~Orazi and M.~Trigiante, 
{\it First Order Description of D=4 static Black Holes and the Hamilton-Jacobi equation,}
[arXiv:hep-th/0905.3938].

\bibitem{Cerchiai:2009pi} B.~L.~Cerchiai, S.~Ferrara, A.~Marrani and B.~Zumino, 
{\it Duality, Entropy and ADM Mass in Supergravity,}
Phys.\ Rev.\  D {\bf 79} (2009) 125010 [arXiv:0902.3973].
%%CITATION = PHRVA,D79,125010;%%

\bibitem{Ceresole:2009id}
  A.~Ceresole, S.~Ferrara and A.~Gnecchi,
  {\it 5d/4d U-dualities and N=8 black holes,}
  Phys.\ Rev.\  D {\bf 80} (2009) 125033
  [arXiv:0908.1069 [hep-th]].
  %%CITATION = PHRVA,D80,125033;%%
%\cite{Ceresole:2009jc}

\bibitem{Ceresole:2009jc}
  A.~Ceresole, S.~Ferrara, A.~Gnecchi and A.~Marrani,
  {\it More on N=8 Attractors,}
  Phys.\ Rev.\  D {\bf 80} (2009) 045020
  [arXiv:0904.4506 [hep-th]].
  %%CITATION = PHRVA,D80,045020;%%

%\cite{Ferrara:2007tu}
\bibitem{Ferrara:2007tu}
  S.~Ferrara and A.~Marrani,
  {\it On the Moduli Space of non-BPS Attractors for N=2 Symmetric Manifolds},
  Phys.\ Lett.\  B {\bf 652} (2007) 111
  [arXiv:0706.1667 [hep-th]].
  %%CITATION = PHLTA,B652,111;%%


\bibitem{Andrianopoli:2010bj}
  L.~Andrianopoli, R.~D'Auria, S.~Ferrara and M.~Trigiante,
  {\it Fake Superpotential for Large and Small Extremal Black Holes},
  arXiv:1002.4340 [hep-th].
  %%CITATION = ARXIV:1002.4340;%%  

\bibitem{new} 
A. ~Ceresole, S.~Ferrara and A.~Marrani, {\it Small N=2 Extremal Black Holes in Special Geometry}
[arXiv:1006.2007[hep-th]] 

%\cite{Cerchiai:2010xv}
\bibitem{Cerchiai:2010xv}
  B.~L.~Cerchiai, S.~Ferrara, A.~Marrani and B.~Zumino,
  {\it Charge Orbits of Extremal Black Holes in Five Dimensional Supergravity},
  arXiv:1006.3101 [hep-th].
  %%CITATION = ARXIV:1006.3101;%%
  
  %\cite{Borsten:2010aa}
\bibitem{Borsten:2010aa}
  L.~Borsten, D.~Dahanayake, M.~J.~Duff, S.~Ferrara, A.~Marrani and W.~Rubens,
  {\it Observations on Integral and Continuous U-duality Orbits in N=8
  Supergravity},
  arXiv:1002.4223 [hep-th].
  %%CITATION = ARXIV:1002.4223;%%

%\cite{Strominger:1996sh}
\bibitem{SV}
  A.~Strominger and C.~Vafa,
  {\it Microscopic Origin of the Bekenstein-Hawking Entropy},
  Phys.\ Lett.\  B {\bf 379}, 99 (1996)
  [arXiv:hep-th/9601029].
  %%CITATION = PHLTA,B379,99;%%

%\cite{Ooguri:2004zv}
\bibitem{OSV}
  H.~Ooguri, A.~Strominger and C.~Vafa,
  {\it  Black hole attractors and the topological string},
  Phys.\ Rev.\  D {\bf 70} (2004) 106007
  [arXiv:hep-th/0405146].
  %%CITATION = PHRVA,D70,106007;%%

\bibitem{DeWit} 
%\cite{Lopes Cardoso:2006bg}
%\bibitem{Lopes Cardoso:2006bg}
  G.~Lopes Cardoso, B.~de Wit, J.~Kappeli and T.~Mohaupt,
  {\it Black hole partition functions and duality}
  JHEP {\bf 0603} (2006) 074
  [arXiv:hep-th/0601108].
  %%CITATION = JHEPA,0603,074;%%
%\cite{Lopes Cardoso:2004xf}
%\bibitem{Lopes Cardoso:2004xf}
  G.~Lopes Cardoso, B.~de Wit, J.~Kappeli and T.~Mohaupt,
  {\it Asymptotic degeneracy of dyonic N = 4 string states and black hole
  entropy},
  JHEP {\bf 0412} (2004) 075
  [arXiv:hep-th/0412287].
  %%CITATION = JHEPA,0412,075;%%

%\cite{Dabholkar:2004dq}
\bibitem{Maloney}
  A.~Dabholkar, R.~Kallosh and A.~Maloney,
  {\it A stringy cloak for a classical singularity},
  JHEP {\bf 0412} (2004) 059
  [arXiv:hep-th/0410076].
  %%CITATION = JHEPA,0412,059;%%

%\cite{Maldacena:1997de}
\bibitem{Maldacena:1997de}
  J.~M.~Maldacena, A.~Strominger and E.~Witten,
  {\it Black hole entropy in M-theory,}
  JHEP {\bf 9712}, 002 (1997)
  [arXiv:hep-th/9711053].
  %%CITATION = JHEPA,9712,002;%%
\bibitem{Sen} A. Sen,
{\it Black Hole Entropy Function, Attractors and Precision Counting of Microstates}
 Gen.Rel.Grav.{\bf 40}  (2008) 2249-2431, 
e-Print: arXiv:0708.1270 [hep-th]


%\cite{Dabholkar:2006zz}
\bibitem{Dabholkar}
  A.~Dabholkar,
 {\it Black hole entropy in string theory: Going beyond Bekenstein and Hawking},
  Int.\ J.\ Mod.\ Phys.\  D {\bf 15}, 1561 (2006);
  %%CITATION = IMPAE,D15,1561;%%  
 % \bibitem{Dabholkar:2005by}
  A.~Dabholkar, F.~Denef, G.~W.~Moore and B.~Pioline,
  {\it  Exact and Asymptotic Degeneracies of Small Black Holes},
  JHEP {\bf 0508}, 021 (2005)
  [arXiv:hep-th/0502157].
  %%CITATION = JHEPA,0508,021;%%
  
  %\cite{Dabholkar:2005dt}
\bibitem{Dabholkar:2005dt}
  A.~Dabholkar, F.~Denef, G.~W.~Moore and B.~Pioline,
  {\it  Precision counting of small black holes},
  JHEP {\bf 0510}, 096 (2005)
  [arXiv:hep-th/0507014].
  %%CITATION = JHEPA,0510,096;%%
%\cite{Dabholkar:2005by}

%\cite{Shih:2005qf}
\bibitem{Shih:2005qf}
  D.~Shih, A.~Strominger and X.~Yin,
  {\it Counting dyons in N = 8 string theory},
  JHEP {\bf 0606}, 037 (2006)
  [arXiv:hep-th/0506151].

  %\cite{Denef:2007vg}
\bibitem{Denef:2007vg}
  F.~Denef and G.~W.~Moore,
  {\it Split states, entropy enigmas, holes and halos},
  arXiv:hep-th/0702146.
  %%CITATION = HEP-TH/0702146;%%

%\cite{Denef:2007yt}
\bibitem{Denef:2007yt}
  F.~Denef, D.~Gaiotto, A.~Strominger, D.~Van den Bleeken and X.~Yin,
  {\it  Black hole deconstruction},
  arXiv:hep-th/0703252.
  %%CITATION = HEP-TH/0703252;%%

%\cite{Sen:2008ht}
\bibitem{Sen:2008ht}
  A.~Sen,
  {\it Wall Crossing Formula for N=4 Dyons: A Macroscopic Derivation},
  JHEP {\bf 0807}, 078 (2008)
  [arXiv:0803.3857 [hep-th]];
  %%CITATION = JHEPA,0807,078;%%
  %\cite{Sen:2008ta}
%\bibitem{Sen:2008ta}
  %A.~Sen,
  {\it N=8 Dyon Partition Function and Walls of Marginal Stability},
  JHEP {\bf 0807}, 118 (2008)
  [arXiv:0803.1014 [hep-th]].
  %%CITATION = JHEPA,0807,118;%%

%\cite{Gaiotto:2009hg}
\bibitem{Gaiotto:2009hg}
  D.~Gaiotto, G.~W.~Moore and A.~Neitzke,
  {\it Wall-crossing, Hitchin Systems, and the WKB Approximation},
  arXiv:0907.3987 [hep-th];
  %%CITATION = ARXIV:0907.3987;%%
%\cite{Jafferis:2008uf}
%\bibitem{Jafferis:2008uf}
  D.~L.~Jafferis and G.~W.~Moore,
  {\it Wall crossing in local Calabi Yau manifolds},
  arXiv:0810.4909 [hep-th].
  %%CITATION = ARXIV:0810.4909;%%

\bibitem{Gun} 
%\cite{Gunaydin:2007bg}
%\bibitem{Gunaydin:2007bg}
  M.~Gunaydin, A.~Neitzke, B.~Pioline and A.~Waldron,
  {\it Quantum Attractor Flows},
  JHEP {\bf 0709}, 056 (2007)
  [arXiv:0707.0267 [hep-th]].
  %%CITATION = JHEPA,0709,056;%%


\bibitem{geo} 
%\cite{Bergshoeff:2008be}
%\bibitem{Bergshoeff:2008be}
  E.~Bergshoeff, W.~Chemissany, A.~Ploegh, M.~Trigiante and T.~Van Riet,
  {\it Generating Geodesic Flows and Supergravity Solutions}
  Nucl.\ Phys.\  B {\bf 812}, 343 (2009)
  [arXiv:0806.2310 [hep-th]].
  %%CITATION = NUPHA,B812,343;%%
  %\cite{Bossard:2009at}
\bibitem{Bossard:2009at}
  G.~Bossard, H.~Nicolai and K.~S.~Stelle,
  {\it Universal BPS structure of stationary supergravity solutions},
  JHEP {\bf 0907}, 003 (2009)
  [arXiv:0902.4438 [hep-th]].
  %%CITATION = JHEPA,0907,003;%%
%\cite{Guica:2008mu}
\bibitem{Guica:2008mu}
  M.~Guica, T.~Hartman, W.~Song and A.~Strominger,
  {\it The Kerr/CFT Correspondence},
  Phys.\ Rev.\  D {\bf 80}, 124008 (2009)
  [arXiv:0809.4266 [hep-th]].
  %%CITATION = PHRVA,D80,124008;%%


\bibitem{DuffQIT}
%\cite{Borsten:2008wd}
%\bibitem{Borsten:2008wd}
  L.~Borsten, D.~Dahanayake, M.~J.~Duff, H.~Ebrahim and W.~Rubens,
  {\it Black Holes, Qubits and Octonions},
  Phys.\ Rept.\  {\bf 471}, 113 (2009)
  [arXiv:0809.4685 [hep-th]];
  %%CITATION = PRPLC,471,113;%%
  %\cite{Duff:2010zz}
%\bibitem{Duff:2010zz}
  M.~Duff,
  {\it  Black Holes And Qubits},
  CERN Cour.\  {\bf 50N4}, 13 (2010).
  %%CITATION = CECOA,50N4,13;%%

%\cite{Bena:2009ev}
\bibitem{Bena:2009ev}
  I.~Bena, G.~Dall'Agata, S.~Giusto, C.~Ruef and N.~P.~Warner,
  {\it Non-BPS Black Rings and Black Holes in Taub-NUT},
  JHEP {\bf 0906} (2009) 015
  [arXiv:0902.4526 [hep-th]].
  %%CITATION = JHEPA,0906,015;%%

 \bibitem{Gaiotto:2005xt}
  D.~Gaiotto, A.~Strominger and X.~Yin,
  {\it 5D black rings and 4D black holes},
  JHEP {\bf 0602}, 023 (2006)
  [arXiv:hep-th/0504126];
  %%CITATION = JHEPA,0602,023;%%
%\cite{Gaiotto:2005gf
%\bibitem{Gaiotto:2005gf}
  D.~Gaiotto, A.~Strominger and X.~Yin,
  {\it New Connections Between 4D and 5D Black Holes},
  JHEP {\bf 0602}, 024 (2006)
  [arXiv:hep-th/0503217].
  %%CITATION = JHEPA,0602,024;%%

\bibitem{GOS}
%\bibitem{Green:2007zzb}
  M.~B.~Green, H.~Ooguri and J.~H.~Schwarz,
  {\it Decoupling Supergravity from the Superstring},
  Phys.\ Rev.\ Lett.\  {\bf 99}, 041601 (2007)
  [arXiv:0704.0777 [hep-th]].

\bibitem{BFK}
%\bibitem{Bianchi:2009mj}
  M.~Bianchi, S.~Ferrara and R.~Kallosh,
  {\it Observations on Arithmetic Invariants and U-Duality Orbits in N =8 Supergravity},
  JHEP {\bf 1003}, 081 (2010)
  [arXiv:0912.0057 [hep-th]];
%\bibitem{Bianchi:2009wj}
 % M.~Bianchi, S.~Ferrara and R.~Kallosh,
  {\it Perturbative and Non-perturbative N =8 Supergravity},
  arXiv:0910.3674 [hep-th].


\end{thebibliography}
\end{document}